\documentclass[12pt]{iopart}
\usepackage{iopams}  
\usepackage{graphics, epsfig}

\begin{document}
\jl{32}

\title[Violation of a Bell-like Inequality in Neutron-Optical
Experiments] {Violation of a Bell-like Inequality in
Neutron Optical Experiments: Quantum contextuality }

\author{Yuji Hasegawa\dag\footnote[3]{To
whom correspondence should be addressed.}, 
Rudolf Loidl\ddag\, Gerald Badurek\dag,
Matthias Baron\ddag\footnote[4]{Present address: Atominstitut der
\"{O}sterreichischen Universit\"{a}ten, Stadionallee 2, A-1020 Wien,
Austria} and  Helmut Rauch\dag}

\address{\dag\ Atominstitut der \"{O}sterreichischen Universit\"{a}ten,
Stadionallee 2, A-1020 Wien, Austria}

\address{\ddag\ Institute Laue Langevin, B. P. 156, F-38042 Grenoble Cedex
9, France}

\begin{abstract}
We report a single-neutron optical experiment to demonstrate the
violation of a Bell-like inequality. Entanglement is achieved not between particles, but between the degrees of
freedom, in this case, for a single-particle. The spin-{\small 1/2}
property of neutrons are utilized. The total wave function of the neutron
is described in a tensor product Hilbert space. A Bell-like inequality is
derived not by a non-locality but by a contextuality. Joint measurements
of the spinor and the path properties lead to the violation of a
Bell-like inequality. Manipulation of the wavefunction in one Hilbert
space influences the result of the measurement in the other Hilbert
space. A discussion is given on the quantum contextuality and an
entanglement-induced correlation in our experiment.

\end{abstract}



\maketitle

\section{Introduction}

In a classic paper, Einstein, Podolsky, and Rosen (EPR) suggested the
incompleteness of quantum theory by considering the reality given by the
wave function of two physical quantities \cite{EINSTEIN}. Their 'paradox'
was moved forward to additional supplemented variables, towards a theory
intending to reestablish a quantum mechanics with both causality and
locality \cite{BOHM}. It was Bell who introduced an inequality to show the
inconsistency of the hidden-variable theory and the statistical predictions
of quantum theory \cite{BELL}. His inequality was reformulated in
various ways to adapt to real experimental tests of local hidden-variable
theories. The most useful form may be so-called CHSH inequalities
(Clauser-Horne-Shimony-Holt inequalities) \cite{CHSH}. Bell's
inequalities together with the EPR-paradox have aroused considerable
interests for many decades \cite{CLAUSER, MANN, AFRAIT}. The contradiction
inherent in such non-local phenomena with our common experience can be
intuitively understood, because one usually considers the result of any
specific measurement independent of what is to be measured at a distance.
Within quantum terminology, this non-locality can be interpreted as a
consequence of the correlation between commuting observables due to the
different position where the measurements are done. Thus, a more general
concept, i.e., contextuality, compared to non-locality can be introduced
to describe other striking phenomena predicted by quantum theory
\cite{KOCHENS, MERMIN1, MERMIN2}. Quantum contextuality is defined as
follows: the result of a measurement of $\hat{\rm A}$ {\it depends} on
another measurement on observable $\hat{\rm B}$, although these two
observables commute with each other. It should be emphasized here that
the property of locality is a special case of non-contextuality.

Except for the experiment with protons \cite{LAMEHI}, most experimental
tests of Bell's inequalities have been performed with correlated photon
pairs \cite{FREEDMAN, ASPECT, OU, KWIAT}. In correlated photon
experiments, two photons are emitted in a J=0 $\rightarrow$ J=1
$\rightarrow$ J=0 atomic cascade or a parametric down-conversion. Thus,
the state of the emitted photon pairs is described by a polarization
entangled state. Since the wave function of each particle belongs to
a different Hilbert space, an observable, e.g., the polarization of one
photon, commutes with those for the other photon. Thus, no correlation
between pairs is expected in general. Entanglement between pair photons,
nevertheless, leads to remarkable correlations in polarization
measurements.

Neutron optical experiments, based on interference of matter waves, have
provided elegant demonstrations of the effects related to the fundamental
aspects of quantum physics. In particular, neutron interferometer
experiments have aroused affection for more than two decades  \cite{BONSE,
KLEIN, BADUREK, RAUCHWERNER}.  Among them, those with a polarized
incident beam served as an ideal tool to investigate properties of a
spin-{\small 1/2} system \cite{POLARIZED}. 

Here, we report an experiment of single-neutron interferometry to show the
violation of a Bell-like inequality. In contrast of the conventional
experiments with entangled particle pairs, the entanglement in our case is
accomplished between different degrees of freedom of a single particle,
i.e., we consider the entanglement between the spinor part and the spatial
part of the wave function. Spin-{\small 1/2} particles, like neutrons, are
described in a tensor product Hilbert space, $\mit H\rm ={\mit
H}_{\mit 1}\otimes {\rm \mit H}_{\rm \mit 2}$ where ${\mit H}_{\mit 1}$
and ${\mit H}_{\mit 2}$ are disconnected Hilbert spaces corresponding to
the spinor and the spatial wave function respectively. Observables of the
spinor part commute with those of the spatial part, and this justifies
the derivation of a Bell-like inequality equivalent to the rejection of
the hypothesis of local realism \cite{BASU}. The experiment consists of
joint measurements of commuting observables of single neutrons in an
appropriately prepared nonfactorisable state. A brief report of the
experimental results has been published previously
\cite{HASEG}.

\section{Theory}

Most experimental tests of the violation of Bell's inequalities have been
made with correlated photon pairs. There, a source was used to emit
entangled photon pairs in a Bell-state, for instance, 

\begin{equation}
\left|{{\Psi}_{ab}}\right\rangle=
{\left|{H}\right\rangle }_{a}\otimes
{\left|{V}\right\rangle}_{b}+
{\left|{V }\right\rangle}_{a}\otimes
{\left|{H}\right\rangle} _{b} ,
\end{equation}

\noindent
where $\left|{H }\right\rangle$ and $\left|{V }\right\rangle$ 
denote the horizontally and the vertically polarized states respectively.
$\left|{{\Psi }_{ab}}\right\rangle$ represents an entangled state, which
shows a correlation and/or an anticorrelation. Here, CHSH inequalities are
summarized as follows \cite{ASPECT2}:

\begin{equation}
 -2\ \le \rm \ S\ \le \rm \ 2 ,
\end{equation}

\noindent
with 

\noindent
\begin{equation}
S=E\left({\vec{{a}_{1}},\vec{{b}_{1}}}\right)
-E\left({\vec{{a}_{1}},\vec{{b}_{2}}}\right)
+E\left({\vec{{a}_{2}},\vec{{b}_{1}}}\right)
+E\left({\vec{{a}_{2}},\vec{{b}_{2}}}\right).   
\end{equation}

\noindent
where $E\left({\vec{{a}_{\mit j}},\vec{{b}_{\mit k}}}\right)$ represents
the expectation value of finding one photon {\it a} in polarization
oriented to $\vec{{a}_{\mit j}}$ and the other {\it b} in polarization
oriented to $\vec{{b}_{\mit k}}$. In real experiments, this polarization
correlation coefficient is given by coincidence counts N, 

\begin{equation}
E\left({\vec{a},\vec{b}}\right)=
{\frac{{N}_{++}\left({\vec{a},\vec{b}}\right)
+{N}_{--}\left({\vec{a},\vec{b}}\right)
-{N}_{+-}\left({\vec{a},\vec{b}}\right)
-{N}_{-+}\left({\vec{a},\vec{b}}\right)}
{{N}_{++}\left({\vec{a},\vec{b}}\right)
+{N}_{--}\left({\vec{a},\vec{b}}\right)
+{N}_{+-}\left({\vec{a},\vec{b}}\right)
+{N}_{-+}\left({\vec{a},\vec{b}}\right)}} .
\end{equation}

\noindent
The signs of "$+$" and "$-$" denote the two outputs of the two-channel
polarization analyzers for the parallel and the (orthogonal) perpendicular
polarization.

Let us consider the principle of the experiment which demonstrates the
violation of a Bell-like inequality in a single-neutron interferometry.
Figure 1 shows the experimental setup. This setup is
similar to that of a previous neutron interferometry for the spin
superposition experiment \cite{SUMMI}. In a polarized neutron
interferometer experiment, the wave function of each neutron is
described in a tensor product Hilbert space by the product of a spinor
and a spatial wave function.  This wave function represents the
entanglement of the spinor part and the spatial part. The normalized
total wave function $\left|{\Psi }\right\rangle$ is given by

\begin{equation}
\left|{\Psi }\right\rangle ={\frac{1}{\sqrt
{2}}}\left({\left|{\downarrow }\right\rangle\otimes \left|{\rm
I}\right\rangle\rm +\left|{\uparrow }\right\rangle\otimes \left|{\rm
II}\right\rangle}\right) .
\end{equation}

\noindent
Here, $\left|{\uparrow }\right\rangle$ and $\left|{\downarrow
}\right\rangle$ denote the up- and down-spin states, and
$\left|{\rm I}\right\rangle$ and $\left|{\rm II}\right\rangle$ the two beam
paths in the interferometer. We introduce two operators projecting the
spin part into orthogonal spin states, ${\frac{\mit
1}{\sqrt{2}}}\left({\left|{\uparrow }\right\rangle\pm \rm {e}^{\mit i\rm
\alpha }\left|{\downarrow }\right\rangle}\right)$, which are given by 

\begin{equation}
{\hat{\mit P}}_{\alpha \rm ;\pm \rm 1}^{s}\mit
={\frac{1}{2}}\left({\left|{\rm \uparrow }\right\rangle\rm \pm \rm 
{e}^{\mit i\rm \alpha }\left|{\downarrow
}\right\rangle}\right)\left({\left\langle{\rm
\uparrow }\right|\rm \pm \rm {e}^{-\mit i\rm \alpha
}\left\langle{\downarrow }\right|}\right) .
\end{equation} 

In the same manner, two other projection operators for the path
into orthogonal states,
${\frac{\mit 1}{\sqrt{2}}}\left({\left|{I}\right\rangle\pm \rm {e}^{\mit
i\chi}\left|{II}\right\rangle}\right)$ are defined as,  

\begin{equation}
{\hat{\mit P}}_{\chi \rm ;\pm \rm
1}^{p}\mit={\frac{1}{2}}\left({\left|{\rm I}\right\rangle\rm \pm \rm
{e}^{\mit i\rm
\chi }\left|{II}\right\rangle}\right)\left({\left\langle{\rm I}\right|\rm
\pm \rm {e}^{-\mit i\rm \chi }\left\langle{II}\right|}\right) .
\end{equation} 

\noindent
Here, parameters $\alpha$ and $\chi$ describe the spinor rotation
and the phase shift in the experiments.

The expectation value for a joint measurement of the spin
state ${\frac{\mit 1}{\sqrt {2}}}\left({\left|{\uparrow
}\right\rangle+{e}^{\mit i\rm \alpha }\left|{\downarrow
}\right\rangle}\right)$ and the path ${\frac{\mit 1}{\sqrt
{2}}}\left({\left|{I}\right\rangle+{e}^{\mit i\chi
}\left|{II}\right\rangle}\right)$ is calculated to be

\begin{eqnarray}
\label{eq:2.8}
E'\left({\alpha \rm ,\chi }\right)&=\left\langle{\Psi}\right|
{\hat{P}}^{s}\left({\alpha }\right)\cdot {\hat{\rm P}}^{\rm
p}\left({\chi }\right)\rm \left|{\Psi }\right\rangle \nonumber\\
&=\left\langle{\Psi }\right|\left[{\left({+1}\right){\cdot \hat{\rm \mit
P}}_{\alpha \rm ;+1}^{s}+\left({-1}\right)\cdot {\hat{\rm \mit P}}_{\alpha
\rm ;-1}^{\rm s}}\right] \nonumber \\ &  \ \ \ \ \ \ \times
{\left[{\left({+1}\right){\cdot \hat{\rm \mit P}}_{\chi \rm
;+1}^{p}+\left({-1}\right)\cdot {\hat{\rm \mit P}}_{\chi \rm ;-1}^{\rm
p}}\right]}^{}\left|{\Psi }\right\rangle,
\end{eqnarray}

\noindent
where ${\hat{P}}^{s}\left({\alpha }\right)$ and ${\hat{\rm P}}^{\rm
p}\left({\chi }\right)$ are observables for the spin and the path,
respetively, and are decomposed by the projection operators given by
Eqs. (6) and (7). This expectation value is analogous to that of a joint
measurement, $E\left({\vec{{a}_{\mit j}},\vec{{b}_{\mit k}}}\right)$,  for
correlated-photon-pair experiments \cite{ASPECT2}. It should be emphasized
here that the observables ${\hat{P}}^{s}$ and ${\hat{P}}^{p}$ operate on
the different Hilbert space, then they commuting with each other.

A Bell-like inequality for the single-neutron interferometry uses the
quantity S$'$ which is expressed
with expectation values $E'\left({\alpha \rm ,\chi }\right)$
\cite{BASU}: 

\begin{equation}
 -2\ \le \rm \ S'\ \le \rm \ 2 ,
\end{equation}

\noindent
with 

\noindent
\begin{equation}
S'=E'\left({{\alpha }_{1},{\chi}_{\mit 1}}\right)
-E'\left({{\alpha}_{1},{\chi }_{\mit 2}}\right)
+E'\left({{\alpha}_{2},{\chi }_{\mit1}}\right)
+E'\left({{\alpha}_{2},{\chi }_{\mit 2}}\right) .
\end{equation}

In two-photon correlation experiments, the polarization correlation
coefficient is practically given with coincidence counts N. We consider
here in more detail of the practical deviations of the expectation value
$E'\left({\alpha \rm ,\chi }\right)$. The operators ${\hat{\mit
P}}_{\alpha \rm ;+1}^{s}$ and ${\hat{\mit P}}_{\chi \rm ;+1}^{p}$ can be
realized with the spin-rotator for $\alpha$ and the phase shifter for
$\chi$. In addition,  since
${\hat{\mit P}}_{\alpha \rm ;-1}^{s}={\hat{\mit P}}_{\alpha +\pi \rm
;+1}^{s}$ and ${\hat{\mit P}}_{\chi \rm ;-1}^{p}={\hat{\mit P}}_{\chi
+\pi \rm ;+1}^{p}$, the operators ${\hat{\mit P}}_{\alpha \rm ;-1}^{s}$
and ${\hat{\mit P}}_{\chi \rm ;-1}^{p}$ can also be realized with the
appropriate rotation of the spin-rotator $\alpha+\pi$ and position of the
phase shifter $\chi+\pi$. Therefore, the expectation value
$E'\left({\alpha \rm ,\chi }\right)$ will be practically given by 

\begin{eqnarray}
\fl
\label{eq:2.11}
E'\left({\alpha \rm ,\chi }\right)&=&{\frac
{{N'}_{\small{++}}\left({\alpha \rm ,\chi }\right)
+{N'}_{\small{--}}\left({\alpha \rm ,\chi }\right)
-{N'}_{\small{+-}}\left({\alpha \rm ,\chi }\right)
-{N'}_{\small{-+}}\left({\alpha \rm ,\chi }\right)}
{{N'}_{\small{++}}\left({\alpha \rm ,\chi }\right)
+{N'}_{\small{--}}\left({\alpha \rm ,\chi }\right)
+{N'}_{\small{+-}}\left({\alpha \rm ,\chi }\right)
+{N'}_{\small{-+}}\left({\alpha \rm ,\chi }\right)}} \nonumber\\
&=&{\frac
{{N'}_{\small{++}}\left({\alpha \rm ,\chi }\right)
+{N'}_{\small{++}}\left({\alpha \rm +\pi \rm ,\chi \rm +\pi }\right)
-{N'}_{\small{++}}\left({\alpha \rm ,\chi \rm +\pi }\right)
-{N'}_{\small{++}}\left({\alpha \rm +\pi \rm ,\chi }\right)}
{{N'}_{\small{++}}\left({\alpha \rm ,\chi }\right)
+{N'}_{\small{++}}\left({\alpha \rm +\pi \rm ,\chi \rm +\pi }\right)
+{N'}_{\small{++}}\left({\alpha \rm ,\chi \rm +\pi }\right)
+{N'}_{\small{++}}\left({\alpha \rm +\pi \rm ,\chi }\right)}} , 
\nonumber\\
\end{eqnarray}

\noindent
where ${N'}_{++}\left({{\alpha }_{\mit j},{\chi }_{\mit k}}\right)$
denotes the count rate with the spin-rotation of ${\alpha }_{\mit j}$ and
the phase shift of ${\chi }_{\mit k}$, which is described by 

\begin{equation}
{N'}_{++}\left({{\alpha }_{\mit j},{\chi }_{\mit
k}}\right)=\left\langle{\Psi }\right|{\hat{\mit P}}_{{\alpha }_{\mit
j};+1}^{s}{\cdot \hat{\rm \mit P}}_{{\chi }_{\mit k};+1}^{p}\left|{\Psi
}\right\rangle .
\end{equation}

Relations between the coincidence counts, N, in the two-photon
correlation experiments and the count rates, N$'$, in the single-neutron
interferometer experiment are summarized in the following:

\begin{eqnarray}
\eqalign{\label{eq:2.13}
{N}_{++}\left({\vec{a},\vec{b}}\right)&: {N'}_{++}\left({\alpha \rm
,\chi }\right)\\
{N}_{--}\left({\vec{a},\vec{b}}\right)&: {N'}_{--}\left({\alpha \rm ,\chi
}\right)={N'}_{++}\left({\alpha \rm +\pi \rm ,\chi \rm +\pi }\right)\\
{N}_{+-}\left({\vec{a},\vec{b}}\right)&: {N'}_{+-}\left({\alpha \rm ,\chi
}\right)={N'}_{++}\left({\alpha \rm ,\chi \rm +\pi }\right)\\ 
{N}_{-+}\left({\vec{a},\vec{b}}\right)&: {N'}_{-+}\left({\alpha \rm ,\chi
}\right)={N'}_{++}\left({\alpha \rm +\pi \rm ,\chi}\right).}
\end{eqnarray}

\noindent
Equations (\ref{eq:2.13}) account for the possibility to measure the count
rates N$'$ not simultaneously with different detectors but successively
with one detector. The appropriate observables are tuned by the
spin-rotator and the phase shifter. In this case, a fair-sampling
hypothesis is required to justify obtaining expectation values from such
successive measurements.

	Quantum theory predicts a sinusoidal behaviour for the count rate
${{N'}_{++}}^{qm}\left({\alpha \rm ,\chi }\right)
={\frac{1}{2}}\left\{{1+\cos\left({\alpha \rm +\chi }\right)}\right\}$.
The same behaviour is also derived for the expectation value
$E'\left({\alpha \rm ,\chi }\right)=\cos\left({\alpha \rm +\chi }\right)$.
These functions will lead to a violation of the Bell-like inequality for
various sets of polarization analysis and phase shift. The maximum
violation is expected, for instance, for the following set:
$\alpha_1={\small \pi/2}$, $\alpha_2=0$, $\chi_1=-{\small \pi/4}$, and
$\chi_2={\small \pi/4}$ to show
$S'=2\sqrt {2}=2.82>2 $.

So far we have described the experiment in terms of perfect
implementation. In the actual experiment, however, perfect sinusoidal
dependence of N$'$ cannot be established due to unavoidable component
misalignments, imperfect quality of polarization/interference, etc.,
which is characterized by contrasts of the oscillations. When the
visibility of the oscillation due to the spin-rotation and/or the phase
shift is reduced, it is expected that the theoretical predictions
for N$'$ and E$'$ reduce proportionally, thus S$'$ getting smaller by the
same factor. The experiment to demonstrate the violation of the Bell-like
inequality requires a final visibility higher than $\sqrt {2}/2=70.7\%$.
The interfering H-beam does not satisfy this condition in practical
neutron interferometer experiments. Thus, our experiment utilizes only the
O-beam, which, in principle, has 100$\%$ visibility. The count rates
${N'}_{\pm \ \pm }\left({\alpha \rm ,\chi }\right)$  are measured not
simultaneously but successively with appropriate spin-rotations and phase
shifts. 

\section{Experiments}

Our experiment consists of three stages: preparation, manipulation, and
detection. The preparation was achieved by the use of a spin-turner after
the polarized beam was split into two paths, producing a Bell state,
$\left|{\Psi }\right\rangle={\frac{1}{\sqrt {2}}}\left({\left|{\rightarrow
}\right\rangle\otimes \left|{\rm I}\right\rangle\rm +\left|{\leftarrow
}\right\rangle\otimes \left|{\rm II}\right\rangle}\right)$. In the
manipulation, the two parameters of $\alpha$ and $\chi$ were adjusted.
And finally, the neutrons with certain properties were detected. We
utilized a polarization analysis on the O-beam to obtain correlation
coefficients.

\begin{figure}[h]
\begin{center}
\vspace{0mm}
\includegraphics[width=10cm]{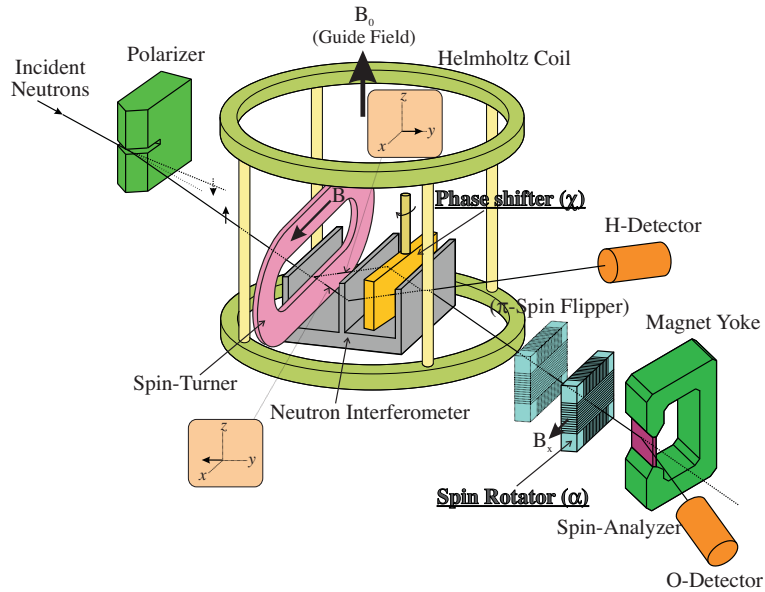}
\caption{ Schematic view of the experimental setup to demonstrate the
violation of a Bell-like inequality in single-neutron interferometry. The
experiment consists of three sages: a preparation of the entangled state
$\left|{\Psi}\right\rangle={\frac{1}{\sqrt {2}}}\left({\left|{\rightarrow
}\right\rangle\otimes\left|{\rm I}\right\rangle\rm +\left|{\leftarrow
}\right\rangle\otimes\left|{\rm II}\right\rangle}\right)$  with the use of
the spin-turner, a manipulation of the two parameters, a phase shift of
$\chi$ and the spinor rotation angle of $\alpha$ together with a
Heusler-analyser, and a detection.}
\end{center}
\label{fig_setup_B}
\end{figure}

	The experiment was carried out at the perfect crystal interferometer beam
line S18 at the high flux reactor of the Institute Laue Langevin
(ILL) \cite{KROUPA}. A schematic view of the experimental setup is
given in Fig. 1. The neutron beam was
monochromatized to have a mean wave length of
$\lambda_{0}=1.92$\AA\  by the use of a Si perfect crystal
monochromator. The incident beam was polarized vertically by
magnetic-prism refractions, then entering a triple-Laue (LLL)
interferometer. This interferometer was adjusted to give a 220 reflection.
A parallel-sided Si plate was used as a phase shifter (varying $\chi$). A
pair of water-cooled Helmholtz coils produced a fairly uniform magnetic
guide field,
$B_{0}\hat{\bf z}$, around the interferometer. A magnetically saturated
Heusler crystal together with a rectangular spin rotator (adjusting
$\alpha$) and a spin flipper, if necessary, enabled the selection of
neutrons with certain polarization directions.

\begin{figure}[h]
\begin{center}
\vspace{0mm}
\includegraphics[width=10.3cm]{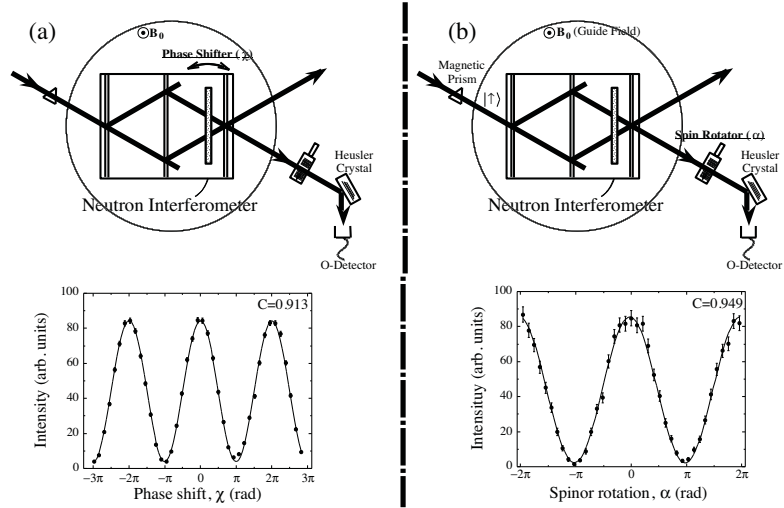}
\caption{
Interference oscillations for two-level systems: (a) for a spatial
Hilbert-space and (b) for a polarization Hilbert-space. The contrasts
over 91$\%$ for (a) and about 95$\%$ for (b) were achieved. This confirmed
the capability of our manipulation for the path and the spinor subsystems.
}
\end{center}
\label{fig_osci_B}
\end{figure}

A crucial optical element in our preparation is a spin turner, which turns
the incident spinor $\left|{\uparrow }\right\rangle$ to
$\left|{\rightarrow }\right\rangle$ for one beam and to $\left|{\leftarrow
}\right\rangle$ for the other. For this procedure, we utilized a
soft-magnetic Mu-metal sheet \cite{RAUCH2}, which gives
considerably high permeability induced by pretty weak magnetic field.
A sheet of 0.5mm thick in an oval ring form was used and two DC-coils
were applied to magnetize this soft-magnetic sheet.

	The important parameters for the manipulation are the relative phase,
$\chi$, between the two beams and the spinor rotation angle, $\alpha$. To
test the capability of our apparatus, we have measured interference
oscillations for two two-level systems in the interferometer: one for a
spatial system, i.e., controlling the path, and the other for a spinor
system, i.e., controlling the spin. Typical oscillations are shown in Fig.
2. One observes sinusoidal oscillations when varying the parameters
$\chi$ for (a) and $\alpha$ for (b). Sufficiently high contrast values
were achieved, which confirmed the capability of our apparatus for
manipulating two subsystems, i.e., the path and the spin, of neutrons in
the interferometer.

\begin{figure}[h]
\begin{center}
\vspace{0mm}
\includegraphics[height=9cm]{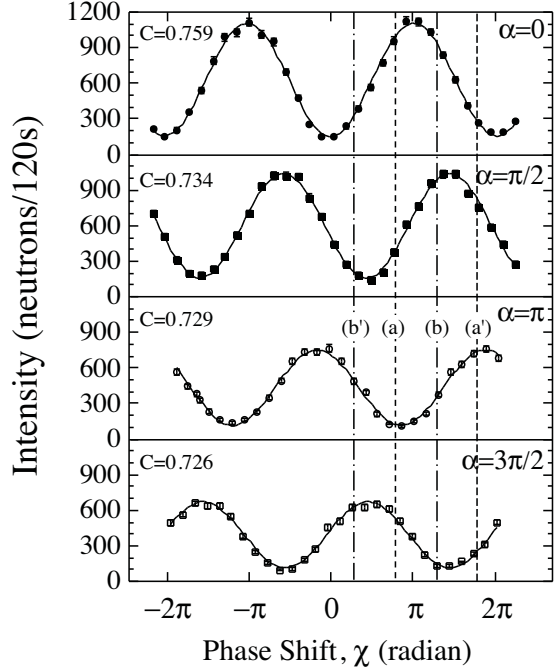}
\caption{
Typical interference oscillations with spinor rotation angle $\alpha=0$,
{\small $\pi$/2}, $\pi$, and {\small 3$\pi$/2}. Contrasts of 76$\%$ for
one and about 73$\%$ for the other three were achieved. Expectation
values, ${E}_{obs}$, were derived from the intensities of appropriate
$\chi$ [on the lines (a) $\chi$=0.79$\pi$ and (a$'$)
$\chi$=1.79$\pi$, or (b)
$\chi$=1.29$\pi$ and (b$'$) $\chi$=0.29$\pi$], where a maximum violation
of the Bell-like inequality is expected. These values exhibited the final
S$'$ value of 2.051$\pm$0.019 $>$ 2: clear violation of the Bell-like
inequality. 
}
\end{center}
\label{fig_result_B}
\end{figure}

A maximum violation of the Bell-like inequality is expected for setting
the spinor rotation angle $\alpha$ at 0, {\small $\pi$/2}, $\pi$, and
{\small 3$\pi$/2}. Typical intensity modulations, obtained by varying the
phase shift $\chi$, are shown in Fig. \ref{fig_result_B}. Contrasts
evidently fell down from those shown in Fig. 2 mainly due to
dephasing/depolarization at the Mu-metal spin-turner. Its gradual
reduction by increasing $\alpha$ is attributed to slight depolarization
by the spinor-rotator and the $\pi$-spin-flipper. We, however, achieved
to obtain enough high contrasts, more than 70.7$\%$, to accomplish the
experiment. Since the Mu-metal spin-turner induces additional relative
phase shift between two-beams in the interferometer, all interference
oscillations are shifted by about $\pi$ in this figure. We took this
shift into account in determining appropriate $\chi$-positions to show the
maximum violation. 

After fitting to sinusoidal dependence by the least squares method, the
expectation values ${E}_{obs}$ were determined using Eq.(\ref{eq:2.11}).
Typical statistical error of ${E}_{obs}$ was ±0.01, obtained from
curves of single measurement. We repeated the same measurements at least
16 times to reduce statistical errors. The final value ${E}_{obs}$ and
its error were evaluated by weighted average of all measurements. So, the
final errors are the sum of systematic and statistical errors. (Main
reason for systematic error was due to phase instability, random drift of
phase, during the measurement.) We obtained ${E}_{obs}\ \left({0,\ 0.79\pi
}\right)$, to be 0.542$\pm$0.007 from the intensities of N$'$(0,\
0.79$\pi$), N$'$($\pi$,\ 0.79$\pi$), N$'$(0,\ 1.79$\pi$), and N$'$($\pi$,\
1.79$\pi$). In the same manner, we obtained
${E}_{obs}\left({0,\ 1.29\pi }\right)=0.4882\pm \rm 0.012$,
${E}_{obs}\left({0.5\pi,\ 0.79\pi }\right)=-0.538\pm \rm 0.006$, and
${E}_{obs}\left({0.5\pi,\ 1.29\pi }\right)=0.438\pm \rm 0.012$. In
evaluating the Bell-like inequality, S$'$ was calculated to be

\begin{eqnarray}
S'&\equiv& \rm E'({\alpha }_{1},{\chi }_{1})+E'({\alpha
}_{1},{\chi}_{2})-E'({\alpha }_{2},{\chi }_{1})+E'({\alpha }_{2},{\chi
}_{2})  \nonumber\\
   &=& 2.051\pm 0.019\ >\ 2 . 
\end{eqnarray}

\noindent
for $\alpha_{1}$, $\alpha_{2}$=0, 0.50$\pi$, and $\chi_{1}$,
$\chi_{2}$=0.79$\pi$, 1.29$\pi$, respectively. This clearly shows a violation
of the Bell-like inequality, which results from the quantum contextuality.

\section{Discussions}

The results given above were obtained by using a neutron detector of more
than 99$\%$ efficiency \cite{KAISER}. In this case, however, a
fair-sampling hypothesis is still required, because of losses in the
interferometer$\--$the second plate of the interferometer is not a
mirror but a beam-splitter$\--$in addition to the fact that the count
rates were obtained successively one after another. It is possible to
introduce the term quantum contextuality in the discussion of our
experiments. The observable for the path and the spin belong to different
Hilbert spaces in our experiments, therefore they commute with each other.
Nevertheless, the expectation value for a joint measurement,
$E'\left({\alpha \rm ,\chi }\right)$ shows correlation between two
observables due to the entanglement between two degrees of freedom of a
single-particle. It is worth noting here that  entanglement is not
limited to different particles but generally applicable to different
subsystems, thus a correlation between variables in single particles
being also expected. General arguments on the entanglement-induced
correlation can be found in the literature
\cite{ENGLERT, BELL2}.

We can interpret the expectation value $E'\left({\alpha \rm ,\chi
}\right)$ in terms of beam polarizations as frequently used in neutron
optics: 'conditional' polarization of 'path' and 'spin' polarizations.
These polarizations vary from $-1$ to +1. The maximum violation of the
Bell-like inequality occurs for a set of four polarization measurements,
three of them resulting in $+\sqrt {2}/2$ and the other in $-\sqrt
{2}/2$. The value E$'$ can not be factorized as 

\begin{eqnarray}
E'\left({\alpha \rm ,\chi }\right) &=& \cos\left({\alpha \rm +\chi
}\right) \nonumber\\  &\ne& \rm \cos\alpha\cdot\rm \cos\chi \rm
={P}^{s}\left({\alpha }\right)\cdot {\rm P}^{\rm p}\left({\chi }\right) ,
\end{eqnarray}

\noindent
although ${\hat{\mit P}}_{\alpha \rm ;\pm \rm 1}^{s}$ and ${\hat{\mit
P}}_{\chi \rm ;\pm \rm 1}^{p}$ commute with each other: a non-factorizable
expectation value E$'$ for the conditional polarization by $\alpha$ for
the spin and $\chi$ for the path. In other words, results of the
measurements for the spin- and the path-polarization are correlated
due to the fact that the subsystems in single neutron are entangled. Such
a correlations involved in our experiments can be used for another
realization of delayed-choice experiment and a new technique of
manipulation, e.g., non-contact control.

\section{Conclusions}

We report an interferometric experiment with a single spin-{\small 1/2}
neutron which demonstrates the violation of a Bell-like inequality. The
wave function of the neutron is described within a tensor product of 
Hilbert spaces, one for the spinor part and the other for the spatial
part of the wave function. Appropriate combinations of spin-rotation
and phase shift lead to the violation of a Bell-like inequality not
with correlated particles but with measurements for single neutrons. A
correlation between subsystems in single particles is induced
by entanglement.

\section*{Acknowledgements}
We appreciate helpful discussions with R.A. Bertlmann (Vienna), I.J.
Cirac (Munich), and D. Home (Calcutta). This work has been supported by
the Austrian Fonds zur F\"{o}derung der Wissenschaftlichen Forschung
(Project No. F1513).

\end{document}